
\magnification=1200
\def\today{\ifcase\month\or
	January\or February\or March\or May\or April\or June\or
	July\or August\or September\or October\or November\or December\fi
	\space\number\day, \number\year}
\rightline {RU--91--34}
\rightline { July 29, 1991}
\vskip 1.5cm
\centerline {\bf A $U(N)$ Gauge Theory in Three Dimensions as an Ensemble of
Surfaces}
\vskip 1cm
\centerline {by}
\vskip 1cm
\centerline{Fran\c cois David\footnote{*}{\it Physique Th\' eorique CNRS.}}
\vskip .2cm
\centerline {\it Service de Physique Th\' eorique\footnote{**} {\it Laboratoire
de l'Institut de Recherche Fondamentale du Commissariat \` a l'Energie
Atomique.}
de Saclay}
\centerline {\it F-91191 Gif-sur-Yvette Cedex}
\vskip .7cm
\centerline {and}
\vskip .7cm
\centerline{Herbert Neuberger}
\vskip .2cm
\centerline{\it Department of Physics and Astronomy}
\centerline{\it Rutgers University}
\centerline{\it Piscataway, NJ 08855--0849}
\vskip 2cm
\centerline{\bf Abstract}
\vskip 1cm
A particular $U(N)$ gauge theory defined on the three dimensional dodecahedral
lattice is shown to correspond to a model of oriented self-avoiding surfaces.
Using large $N$ reduction it is argued that the model is partially soluble in
the planar limit.
\vskip .3cm
\vfill
\eject

The example of three dimensional abelian gauge theory provides the best
understood mechanism for confinement at weak coupling [1-5] and, therefore, a
natural place to look for an equivalent description in terms of a theory of
real strings. As in any gauge theory, a possible place to start from when
looking for string-like excitations is the strong coupling expansion of the
model regularized by discretizing space to a regular  lattice. Roughly, the
closed surfaces that generically appear in the expansion can be thought of as
space-time histories of closed strings. Depending on the nature of the gauge
group these strings may be oriented or not.

The particular case of $U(1)_3$ pure gauge theory with a single plaquette
action of the Villain form [4] is exactly dual to a three dimensional spin
ferromagnet; the spin degrees of freedom are integers and for this reason the
model is sometimes referred to as the $Z$-ferromagnet [5]. The strong coupling
expansion in the gauge formulation is, by duality, related term by term to to
the weak coupling expansion of the $Z$-ferromagnet. The most
commonly studied case is defined on a cubic lattice whose sites we denote by
$x$, $x'$ and whose bonds we represent by their end-points, $<x,x'>$. The
partition function is given by:
$$
Z=\sum_{\{ n(x) \}_{-\infty}^{+\infty}} \exp \{-{{1}\over{2\beta}}
\sum_{<x,x'>} [n(x)-n(x')]^2\}\eqno{(1)}
$$
The surfaces are made out of square plaquettes that live on the cubic lattice
dual to the original one and can be associated with individual spin
configurations $\{ n(x) \} $ by providing walls that separate the original
lattice into connected, non-empty, clusters on which $n$ has a constant value.
Any bond $<x,x'>$ for which $n(x) \ne n(x')$ is cut by a dual
plaquette.

It is well known that in most gauge theories the surface interpretation of
these wall conglomerates can become involved, necessitating some {\sl ad hoc}
definitions, and becoming quite cumbersome [6,7]; to maintain faith in the
existence of a continuum string theory description of these surfaces one must
assume that many of the above complications are irrelevant. It would be nice to
find special forms of the gauge models that avoid some of the complexities
already at the regularized level.

Even in the simple case of $U(1)_3$ on a cubic lattice the surfaces suffer from
complications: plaquettes can be multiply excited in the sense that
$n$ jumps by an amount larger than unity across them and singular lines are
possible where three or more plaquettes join at a common link. We wish to get
rid of these cases and obtain a much cleaner geometrical description of the
surfaces that appear. We first deal with the multiply excited plaquettes by
replacing the action by\footnote{*}{This is a generalization of an action
written down by V. J. Emery and R. Swendsen for an SOS model [8].}
$$
Z=\sum_{\{ n(x) \}_{-\infty}^{+\infty}} \exp \{-{{1}\over{2\beta}}
\sum_{<x,x'>} [n(x)-n(x')]^{2k}\}\eqno{(2)}
$$
and taking the limit $k \rightarrow \infty$; this has the effect of permitting
only jumps of $\pm 1$ across a surface and the two cases can be geometrically
interpreted as being associated with the overall orientation of a closed
connected surface enclosing a given spin cluster (the surfaces are orientable).
This restriction also eliminates cases where three
plaquettes share a common link. However, singular lines where four plaquettes
meet are still possible.  To avoid these configurations we place the
$Z$-ferromagnet on a f.c.c lattice rather than on a cubic one. The geometry of
this lattice is such that its dual has exactly three plaquettes meeting at each
link; thus the bad cases we were left with disappear. The single case we need
to make a slightly {\sl ad hoc} decision for is when surfaces touch at a
vertex: for reasons that will become clear later on we decide not to regard the
touching of two otherwise separated pieces of surface as something that
connects them; in other words, when two surfaces touch at a vertex we view the
vertex as split in two, one vertex for each surface and a very small space open
between them.

We ended up with a model of random self-avoiding orientable surfaces that is
very similar to the system shown to be equivalent to the Ising model on the
f.c.c. lattice in previous work [9]. The difference is that in our case we have
to sum over independent orientations for each connected component of the set of
domain walls. This difference is significant because it enhances the entropy of
configurations made out of many small disconnected bubbles. Due to
self-avoidance a gas of such bubbles will exercise a pressure on a surface
spanning a Wilson loop, keeping it flat, and pushing the deconfinement
transition present in the $Z_2$ gauge theory dual to the Ising model to much
lower temperatures, possibly all the way down to zero temperature.

We now proceed to write down the $U(1)_3$ gauge theory dual to our model:
$$
Z_{dual} = \int_{\{\theta_l \}} \prod_p [1+2g \cos (\sum_{l\in p} \theta_l
)]\eqno{(3)}
$$
Here the $\theta_l$ are angles on the links $l$ and the $p$'s are rhombic
plaquettes on the dodecahedral lattice dual to the f.c.c. lattice. The coupling
$g$ is given by $g=\exp (-{{1}\over{2\beta}})$. This action is the simplest
generalization of the $Z_2$ action written down in ref. [9].

It was shown in the Ising case that $Z_2$ could be replaced by $O(N)$ and, by
appropriately scaling $g$ with $N$, a double expansion in $g$ and $N$ could be
viewed as a sum over surfaces weighted by their total area and by the sum of
the Euler characteristics of their connected components if surface touching at
a site is treated as defined above. In our case the generalization will be to
the gauge group $U(N)$ with
$$
Z_{U(N)} = \int_{\{U_l \}} \prod_p [1+2gN {\rm Re}( Tr ( \prod_{l\in p} U_l ))
]\eqno{(4)}
$$
In terms of surfaces we have:
$$
Z_{U(N)} = \sum_{\{S\}} g^{\sum_S (A(S))}~~ N^{\sum_S \chi (S) }~~ 2^{\sum_S
1}\eqno{(5)}
$$
In the above equation $S$ denotes a connected component of the set of
self-avoiding surfaces $\{S\}$ and $A,\chi$ are functions of $S$ giving the
area in plaquettes and the Euler characteristic.

The new model can be subjected to another duality transformation. The resulting
expression is somewhat complicated by the additional curvature terms. We shall
not write down the explicit expression (it can be found by generalizing
references [9] and [10]); all we wish to stress here is that the new terms have
a coupling $\log (N)$ and are local in the spin variables. Hence, their effect
shouldn't be dramatic for $N$ small enough.

As for practically any gauge theory one can extend the standard arguments [11]
to show that large $N$ factorization will hold to any order in the coupling
$g$. Since the group is now $U(N)$ one can embed in the link
variables the group of lattice translations and achieve Eguchi-Kawai reduction
[12]. Because of the non-exponential structure of the action, quenching  should
not be needed for any value of $g$. Quenching does appear to be necessary in
the usual case when some of the new $U(1)$ symmetries of the reduced model get
spontaneously broken by the attraction between the eigenvalues of the link
matrices overcoming their kinematical repulsion [13]. Here this cannot happen
because the action will have too weak an effect. The additional $U(1)$'s have
to be preserved in order to ensure that closed reduced loops have vanishing
expectations when they correspond to open original loops. The reduced model
will consist of a finite
number of matrices, with an action resembling the action of the original model.
Thus the partition function of the reduced model will be a polynomial in $N$
and $g$. As a result, the purely planar contribution to the free energy per
unit volume, $ {1\over{N^2}} \log (Z_{reduced})$, will vanish.

Let us now describe the reduction of the model in some more detail. The main
new point to realize when generalizing from (hyper)cubic lattices is that
reduction can eliminate only the degrees of freedom that are copies of each
other by pure translations; one has therefore to identify the fundamental set
of lattice points that generate the whole crystal by translations only.

We visualize the f.c.c. lattice as a cubic structure (with no sites yet) to
which we add vertices at the centers of all links and all cubes [14]. Each of
the original cubes can be cut into eight smaller cubes, each of which has four
of its corners occupied and the other four free. Any two adjacent small cubes
are mirror images of each other. The dual lattice is made out of vertices that
sit at the unoccupied corners of the little cubes and at their centers. The
bonds on this lattice connect these new centers to the new corners. Two
adjacent cubes have two new corners in common and together with the two new
centers they build up an elementary rhombic plaquette. The smallest shape
enclosed by the rhombi is a dodecahedron and the dodecahedra fill the space
exactly.

It is clear now that the dodecahedral lattice has at least two kinds of
vertices, one with eight links connected to it (a little cube corner) and
another with only four (a little cube center). What is slightly less obvious is
that there are really two kinds of links with coordination number equal to
four, related to each other by reflection through a plane. These two kinds
cannot be mapped one into the other by a pure translation. We therefore end up
with a reduced model consisting of three vertices and eight oriented links. The
eight links start from a central vertex, C, and are connected in two groups of
four to two additional sites, referred to as L(eft) and R(ight). On each of the
links we have a $U(N)$ matrix or its hermitian conjugate, depending on the
direction we traverse the link. Denoting the C---L(R) four link variables by
$U_{\alpha}$ ($V_{\alpha}$) the partition function of the reduced model
becomes:
$$
Z_{reduced} = \int \prod_{\alpha=1}^{4} \prod_{\beta=1}^{4}\{ dU_{\alpha}
dV_{\beta} [1+2gN{\rm Re} (Tr U_{\alpha} U_{\beta}^{\dag} V_{\alpha}
V_{\beta}^{\dag})]\}\eqno{(6)}
$$
Any loop on the original lattice can, modulo translations, be identified by a
sequence of link traversals after an arbitrary starting point has been picked
on the loop. A C---R(L) link traversal must be followed by  a R(L)---C one
respectively, but a R(L)---C link traversal can be followed either by a C---R
or a C---L one. The sequence of link passages can be taken over to the reduced
lattice. We only need to make sure now that sets of links that would correspond
to a curve with different end points on the original lattice can be
distinguished, even after reduction, from the reduced image of an originally
closed curve. For this we need additional $U(1)$ symmetries in the reduced
model under which only the reduced images of closed curves will give a singlet
after taking the $U(N)$ trace. To ensure that a curve indeed closes there are
three conditions corresponding to the three independent coordinates of the
``end point'' that must be identical with the ``starting point''. Hence we need
three additional $U(1)$'s. The $U(1)$ ``counting'' goes as follows: The reduced
model has five $U(1)$'s,
$$
U_{\alpha} \rightarrow e^{i\phi_{\alpha} +i\psi_u} U_{\alpha};~~~~
V_{\alpha} \rightarrow e^{-i\phi_{\alpha} +i\psi_v} V_{\alpha}.\eqno{(7)}
$$
The original model had two non-gauge $U(1)$ symmetries corresponding to the
multiplication by a phase of all the C---R link variables and by another phase
of all the C---L link variables. These two $U(1)$'s
are obviously present in the reduced model too, leaving $5-2=3$ new $U(1)$'s,
the exact needed number.

Armed with the knowledge that factorization holds, one can now simply replay
the Eguchi-Kawai [12] derivation of the equivalence of the reduced model to the
original one. As we already mentioned, there is no reason to suspect that
the additional $U(1)$'s will break spontaneously and therefore quenching won't
be necessary. In view of the polynomial form of the action it is plausible that
the model is essentially soluble and that explicit expressions for the
expectation values of the traces of many Wilson loop operators can be written
down. We are not going to pursue these matters any further here.

Instead we turn to making several observations about the structure of the
model.

The partition function of the original model wouldn't change if we change the
space the link variables take values in from $U(N)$ to $SU(N)$ as long as $N\ge
4$. Moreover, no change in Wilson loop averages will occur if the Haar
integration measure for each link variable is altered by multiplication by
$\exp [\rho(U_{l})]$ where $\rho$ is a class function also invariant under
multiplication of its argument by an element of the center of the group. A
similar remark holds for the model of ref. [9]. This shows that we have real
sensitivity only to the center of the group, in accordance with one of the more
popular mechanisms for confinement. Note that there is a difference between the
case that the center is strictly $Z_2$ and when the center contains $Z_4$. The
$Z_3$ case seems special and indeed its dual would be a $Z_3$ spin model which,
in three dimensions, in the simple cases, will have no continuous phase
transitions.

There is a non-trivial issue that has to be brought up regarding the expected
importance of the restriction $|n(x)-n(x')|\le 1$ on the magnitude of the jump
between nearest neighbors in the $Z$-ferromagnet. We would like the restriction
to have no dramatic effect when $\beta$ is
very large, in particular not to have a deconfinement transition at a finite
$\beta$.
Superficially it seems that the restriction, if anything, will only aid
confinement because
it helps the $n(x) \rightarrow n(x)+n_0$
symmetry of the dual spin system to stay broken. However, there might be a flaw
in the argument
because the model can also be viewed as a restricted $Z_4$ spin
model.\footnote{*}{Here we
generalize some observations made in ref. [15].} To see this, let us work for
the moment in
a finite volume with free boundary conditions and fix $n(x_{0})$ at some site
$x_0$ to zero.
Consider the set of
``pure gauge fields'' (on the f.c.c. lattice) consisting of the differences
$n(x)-n(x')$
across oriented bonds
and associate to each such link the angle $\theta (x,x') = {{\pi}\over{2}}
[n(x)-n(x')]$. One can
think about these angles as a set of ``pure gauge fields'' for the gauge group
$Z_4$. Setting $\theta(x_{0})=0$ one can construct a unique $Z_4$ spin
configuration that would gauge transform
$\theta (x,x')$ to zero everywhere. The set $\{\theta (x)\}_{\theta( x_0 ) =0}$
is in one to one correspondence with the set $\{n(x)\}_{n( x_0 ) =0}$ if the
angle $\theta (x) $ is not let to
rotate by more
than ninety degrees along any bond. The difference between the restricted and
unrestricted
model can be also seen in another way: in the restricted model averages of
Wilson loops that
carry a charge larger than two vanish exactly.

If one thinks about the model as a model of real surfaces representing boundary
free membranes
in a fluid one may interpret the additional factor of two per connected
component as arising
from the averaging over a degree
of freedom internal to the surface. For example, one could imagine that on
each surface there lives an interacting two dimensional Ising system whose
self-coupling is infinite, but whose degrees of freedom are otherwise decoupled
from the medium.
The surface entropy factor arises from summing over the two possible states of
the magnetization
in each connected component of the membrane.\footnote {**}{This case would
represent a particular
limit of a model studied in ref. [16].} From this point of view one can
generalize the $Z_2$ model
of ref. [9] even further by admitting $p$ states per surface and increasing
thus the entropy factor
to $p$ per connected component. When formulated in terms of bulk spin variables
this model can be
viewed as consisting of spins that can take values on a homogeneous Bethe
lattice of coordination
$p$. When moving across an elementary bond a spin value can at most jump to a
nearest neighbor on
the Bethe lattice. The case we described in more detail in this note
corresponds to $p=2$.

Suppose that the class of models discussed in the present note, as well as more
traditional formulations, all are related to each other by admitting a
continuum limit that is described by a
string theory. Polyakov has conjectured that the three dimensional Ising model
is described in
the critical regime by a fermionic free string theory [1,17]. These two
situations are different:
While the Ising string would describe a system that is known to be completely
described by an
ordinary (but strongly interacting) field theory, the $U(1)$ gauge case
probably admits no
continuum field theoretical description in the limit where the scale is set by
the string
tension (the regularized form of the field theory is more or less a three
dimensional
Sine-Gordon model, hence perturbatively non-renormalizable). There exists a
decorated loop
operator in the $Z_2$ case (at least on the cubic lattice) that obeys a linear
loop equation
(up to self-intersections, and these are not rapidly generated); there exists
no known analogue
in the $U(1)$ case (the Schwinger-Dyson equation for the Wilson loop will
rapidly generate self-intersections). Our present note and the previous paper
on the $Z_2$ case [9] have shown
that the models admit the introduction of a parameter that might be interpreted
as a ``bare''
string coupling constant; the critical properties of the models seem
insensitive to small
variations in this coupling in both cases, indicating that if a ``physical''
string coupling
does make its appearance eventually, it will have an intrinsically determined
value that cannot
be tuned at will.

It would be interesting to formulate precise numerical tests for the
conjectures that either
theory is represented by a self-consistent, complete string theory. Some
attempts in this
direction have been made in references [18]. The simplest approach conceptually
would be to
try to see some sign of Regge behavior, for example by identifying a few low
lying resonances of moderate spin. The $U(1)$ case seems to be under good
control numerically, beyond bulk
properties, as the basic ideas about confinement have been recently
convincingly tested
quantitatively [19], so there is some hope. Since the f.c.c. lattice is a stack
of two
dimensional triangular lattices the identification of states of higher spin
might be easier
here than in the cubic case.

Our main purpose in the present note was to show that three dimensional gauge
theories are an interesting place to look for new understandings of systems of
fluctuating surfaces or of string theories.

\vskip 1cm
\noindent {\bf Acknowledgements.} This research was supported in part by the
DOE under
grant \# DE-FG05-90ER40559. FD would like to thank the theoretical particle
physics group at Rutgers University for hospitality while part of this work
was done. HN would like to thank E. Witten for several useful discussions.

\vfill
\eject

\leftline {\bf References}
\vskip .5cm
\item {[1]} A. M. Polyakov, {\bf Gauge Fields and Strings}, Contemporary
Concepts in Physics, Volume 3, Harwood Academic Publisher, (1987).

\item {[2]} A. M. Polyakov, {\bf Nucl. Phys. B120} (1977) 429.

\item {[3]} M. E. Peskin, {\bf Ann. Phys. (NY) 113} (1978) 122; R. Savit, {\bf
Phys. Rev. Lett. 39} (1977) 55; T. Banks, R. Myerson, J. Kogut, {\bf Nucl.
Phys. B129} (1977) 493.

\item {[4]} J. Villain, {\bf J. Phys. (Paris) 36} (1975) 581.

\item {[5]} M. G\" opfert, G. Mack, {\bf Comm. Math. Phys. 82} (1982) 545.

\item {[6]} J.-M. Drouffe, J.-B. Zuber, {\bf Phys. Rep. 102} (1983) 1.

\item {[7]} I. Kostov, {\bf Nucl. Phys. B265[FS15]} (1986) 223; K. H. O'Brien,
J.-B. Zuber, {\bf Nucl. Phys. B253} (1985) 621.

\item {[8]} V. J. Emery, R. Swendsen, {\bf Phys. Rev. Lett. 39} (1977) 1414.

\item {[9]}  F. David, {\bf Europhys. Lett. 9} (1989) 575.

\item {[10]} T. Hofs\" ass, H. Kleinert, {\bf J. Chem. Phys. 86} (1987) 3565.

\item {[11]} See Appendix B in [6].

\item {[12]} T. Eguchi, H. Kawai, {\bf Phys. Rev. Lett. 48} (1982) 1063.

\item {[13]} G. Bhanot, U. M. Heller, H. Neuberger, {\bf Phys. Lett. 113B}
(1982) 47.

\item {[14]} N. W. Ashcroft, N. D. Mermin, {\bf Solid State Physics}
Hoft-Saunders International Editions, New York, NY (1976).

\item {[15]} E. Domany, D. Mukamel, A. Schwimmer, {\bf J. Phys. A. Math. Gen.
13} (1980) L311.

\item {[16]} S. Leibler, D. Andelman, {\bf J. Physique 48} (1987) 2013.

\item {[17]} A. M. Polyakov, {\bf Phys. Lett. 82B} (1979) 247, {\bf Phys. Lett.
103B} (1981) 211; Vl. S. Dotsenko, A. M. Polyakov, in {\bf Advanced Studies in
Pure Mathematics 16}, Academic Press (1988); Vl. S. Dotsenko, {\bf Nucl. Phys.
B285} (1987) 45; A. R. Kavalov, A. G. Sedrakayan {\bf Nucl. Phys. B285[FS19]}
(1987) 264; A. G. Sedrakayan {\bf LAPP-TH-314/90} Annecy preprint (1990); E.
Fradkin, M. Srednicki, L. Susskind, {\bf Phys. Rev. D21} (1980) 2885; S. Samuel
{\bf J. Math. Phys. 21} (1980) 2806; C. Itzykson, {\bf Nucl. Phys. B210[FS6]}
(1982) 477; P. Orland {\bf Phys. Rev. Lett. 59} (1987) 2393.

\item {[18]} M. Caselle, R. Fiore, F. Gliozzi, P. Provero, S. Vinti, {\bf DFTT
26/90} Torino preprint (1990); R. Schrader {\bf Jour. of Stat. Phys. 40} (1985)
533; G. M\" unster, in {\bf Probabilistic Methods in Quantum Field Theory and
Quantum Gravity} Edited by P. H. Damgaard et al. , Plenum Press, New York,
(1990).

\item {[19]} A. Duncan, R. Mawhinney, {\bf Phys. Rev. D43} (1991) 554; R. J.
Wensley, J. D. Stack, {\bf Phys. Rev. Lett. 63} (1989) 1764; T. Sterling, J.
Greensite, {\bf Nucl. Phys. B220} (1983) 327; T. A. DeGrand, D. Toussaint, {\bf
Phys. Rev. D22} (1980) 2478.

\vfill
\eject
\bye